# Graphene/Li-Ion battery

Narjes Kheirabadi [1]*, Azizollah Shafiekhani [2, 3]**

[1]Department of Physics, IAU, Northern Tehran Branch, Tehran, 1667934783, Iran.

[2]Physics Department, Alzahra University, Vanak, Tehran, 1993893973, Iran.

[3]School of Physics, Institute for Research in Fundamental Sciences (IPM), P.O.Box:19395-5531, Tehran, Iran.

**Abstract**

Density function theory calculations were carried out to clarify storage states of Lithium (Li) ions in graphene clusters. The adsorption energy, spin polarization, charge distribution, electronic gap, surface curvature and dipole momentum were calculated for each cluster. Li-ion adsorbed graphene, doped by one Li atom is spin polarized, so there would be different gaps for different spin polarization in electrons. Calculation results demonstrated that a smaller cluster between each two larger clusters is preferable, because it could improve graphene Li-ion batteries; consequently, the most proper graphene anode structure has been proposed.

To be appear in J. Appl. Phys.

*Corresponding author: Tell:+989163690785, E-mail address: narjeskheirabadi@yahoo.com

** Tell: +989121066692, E-mail address: ashafie@ipm.ir.

## 1. Introduction

High density energy production and storage systems, i.e. batteries, fuel cells, electrochemical capacitors and solar cells, are currently under intensive research and development [1]. Lithium (Li) ion batteries have been widely used in portable electronic devices and regarded as promising devices in the application of electric vehicles. The energy density and performance of Li-ion batteries largely depend on the physical and chemical properties of cathode and anode materials. Typically, both electrodes in a Li-ion battery are intercalation compounds, which as their name implies, store $Li^+$ by inserting them into their crystal structure in a topotactic manner [2].

The state-of-the-art anodes suffer from one or more of these problems: limited Li storage capacity, large irreversible capacity loss, low charge/discharge rate capability, and poor capacity retention upon the charge/discharge cycling, etc. [1].

The theoretical specific capacity of graphite is 372 $mAhg^{-1}$ (by forming intercalation compounds $LiC_6$) [3]. Graphite is the commercial anode material widely used for Li batteries because of its high Coulombic efficiency and better cycle performance [3]. Therefore, the anode used in most Li-ion batteries is based on graphite carbon, which stores up to one $Li^+$ for every six carbon atoms between its graphene layers [4].

Due to the capacity limit of graphite, the energy density of Li-ion battery cannot satisfy the requirements of portable electronic devices. Traditional intercalation-type graphite materials show low Li storage capacity (<372 $mAhg^{-1}$, $LiC_6$) due to limited Li ion storage sites within a $sp^2$ hexagonal carbon structure [2]. To meet the increasing demand for batteries of high-energy density, much effort has been made to explore new anode materials [5–9].

In 2004, graphene was obtained by mechanically separating individual graphene sheets from the strongly bonded layered structure of graphite [10, 11]. This technique yields graphene samples that are virtually free of crystal defects, and consequently have a carrier mobility of up to $2 \times 10^5 \text{ cm}^2\text{V}^{-1}\text{s}^{-1}$ [12, 13].

Graphene is one of the building blocks of all kinds of carbonaceous materials. A monolayer graphene sheet can be spirally wrapped into a single wall carbon nanotube. Multiple graphene sheets can be stacked into graphite with multi-layered graphene sheets which, in turn can be clustered into hard carbons or wrapped into multi-wall carbon nanotubes. Therefore, graphene structure provides an ideal platform for fundamental understanding of Li-C [1].

In contrast, graphene, the unique two-dimensional atom-thick honey-comb structured carbon, has revealed various novel properties. Graphene has the highest intrinsic mechanical strength (1060 Gpa) and thermal conductivity (3000 $\text{Wm}^{-1}\text{k}^{-1}$), in addition to a high surface area (2630 $\text{m}^2\text{g}^{-1}$) and electronic mobility (10000 $\text{cm}^2\text{V}^{-1}\text{s}$) [1].

Because of its well-known high carrier mobility and its inherited two-dimensional shape with monatomic thickness, researchers have attempted to fabricate in-plane graphene electronic devices. And, changes in graphene caused by molecular charge transfer should be explored for different applications [14]. Nevertheless, a method for opening the band gap needs to be established before applying graphene as active elements in such devices. In this regard, cutting the graphene in the form of a Nano-ribbon or a super lattice of the hydrogenated domain has been investigated. In a slightly different context, hydrogen chemisorption states of graphene have attracted researchers by the possibility of non-metal magnetism [15–19].

Owing to its vast surface-to-volume ratio and highly conductive nature, graphene may also bring high Li storage capacity to Li-ion batteries [20, 21]. If the carbon material is transformed from graphite to graphene, the capacity is expected to increase up to 500–1100 $\text{mAhg}^{-1}$. Special characteristics of graphene originate from its non-layered structure where Li atom and ion are stored in both surface and edge regions [22].

During the intercalation process, Li transfers its 2s electrons to the carbon host, where it would be situated between the carbon sheets. High capacity carbon materials could be mainly ascribed to (I) lithium insertion within the ''cavities'' of the material [15], (II) lithium adsorbed on each side of the carbon sheet [23], (III) lithium binding on the so called ''covalent'' site [15], and (IV) lithium binding on hydrogen terminated edges of graphene fragments in carbon-containing materials [24, 25]. It has been proposed that lithium ions can be adsorbed on both sides of the graphene sheets which are arranged like a ''house of cards'' in hard carbons, leading to two layers of lithium for each graphene sheet, with a theoretical capacity of 744 mAhg$^{-1}$ through the formation of Li$_2$C$_6$ [10, 15, and 26]. Recently, large reversible Li storage (540 mAhg$^{-1}$ in the first cycle) in graphene Nano-sheets has been reported [20, 21].

To improve rechargeable battery, the capacity of graphene to store hydrogen is so important, especially when it is doped by Li$^+$ [27-30]. To excel the features of Li ion battery, for example, to shorten the charge time [15] or to achieve higher-energy density, the exact number of adsorbed Li in graphene and the adsorbing energy is significant. So, many attempts have been made to plan a graphene layer to adsorb a higher number of Li atoms or to make a better anode by using other compounds [22].

The interaction between Li$^+$ and a graphite surface has been investigated theoretically by several researchers using lithium-small carbon cluster. Sato et al. proposed a Li$_2$ covalent molecule model where each Li atom is trapped in one benzene ring (referred to as a "covalent" site) and predicted Li storage capacity of 1.116 mAhg$^{-1}$ in disordered carbon [3]. Marquez et al. calculated the binding energy of Li$^+$, and hydrogen terminated cluster model (C$_{32}$H$_{18}$) using the Density

Function Theory (DFT) method and indicated that the Li$^+$ ion is preferentially bound outside the cluster model (i.e. on the edge site) [31]. Based on semi-empirical molecular orbital calculations using a C$_{96}$ planer carbon cluster and Li$^+$, Nakadaira et al. suggested that the edge site is more stable than that of the bulk [32]. The tight binding calculations showed that a flat band, composed of side edge carbon atoms, is located near the Fermi level [33]. Ab initio calculations for the interaction of the lithium atom with graphite model clusters indicated that charge transfer from the Li atom to the graphite cluster is important in large-cluster size [34-37]. But properties of Nano-scrolls and edge effects on graphene need wider studies [14]. The current research considers graphene to adsorb one Li ion on each side. DFT method is used to study the models with different Li locations. Larger clusters have been examined in comparison to the reference research [31]. Moreover, a more accurate calculation method has been applied [32]. Finally, a new structure has been proposed for a Li-ion battery which is an important progress in comparison to finite graphene calculations.

## 2. Clusters and Calculation Method

In the present research work, we have tried to map the interaction of Li and circular graphene with 24 to 96 carbon atoms with hydrogen at the boundary (Coronene (C$_{24}$H$_{12}$), Charcoal (C$_{54}$H$_{18}$) and C$_{96}$H$_{24}$), so 10 different clusters are suggested (FIG.1). Due to the symmetry of these clusters, study of edge effects of graphene-doped Li will be very important in quantum state. The calculation method is DFT using Gaussian03 software [38] with the Beke-Lee-Yang-Parr (B3LYP) exchange-correlation hybrid function [39,40] and 6-31g* basis set to study relative energy, band gap, density of states (DOS), surface shape, dipole momentum, and electrical polarization of each cluster. All of DOS diagrams and gap information are extracted by Gauss Sum software [41].

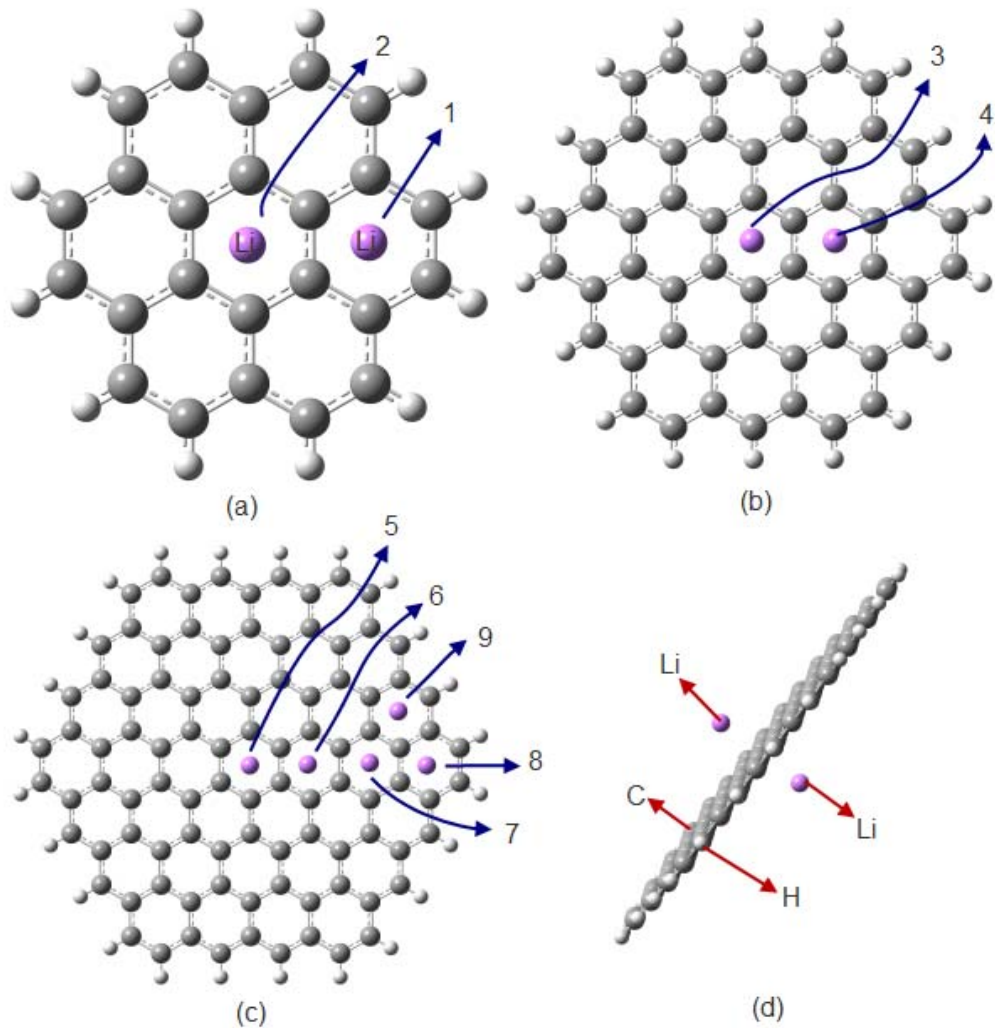

FIG.1. Clusters (a) to (c) are disordered with a Li in a specific place, demonstrated on the figure by cluster name (a) Two different $LiC_{24}H_{12}$ clusters (clusters number 1 and 2) (b) Two distinct $LiC_{54}H_{18}$ clusters (clusters number 3 and 4) (c) Five dissimilar Li $C_{92}H_{24}$ clusters (cluster number 5 through 9) (d) $Li_2C_{92}H_{24}$, cluster number 10 which has two Li on each side.

## 3. Results

In the first step, the adsorbing energy for each cluster has been calculated using optimization energy corresponding to the following equation: $X + NLi \rightarrow Li_N X$

In this equation, X is the sign of the cluster, and N is the number of adsorbed lithium atoms. So, the adsorption energy is the difference between two sides of equation energies. If the optimization energy of X value is assumed as in Ref [42], and the optimization energy for the right side of the equation is as described in Table1, and Li optimization energy is -203.81eV, then it is possible to calculate adsorption energy. The results of the above-mentioned calculations have been demonstrated in the fourth column of the Table1.

The adsorption energy depends on the size of each cluster, so relative energy would change as Table 1: $E\ ad\ _{LiC_{92}H_{24}} < E\ ad\ _{Li\ C_{54}H_{18}} < E\ ad\ _{Li\ C_{24}H_{12}}$.

If the cluster 10 adsorption energy corresponding to the cluster number 5 increases, absorption of Li would decrease the amount of energy [42], and if the number of captivated Li increases, cluster relative stability would also increase; the relative stabilities of different (unstrained) PAH's often are based on the resonance energies or the aromatic stabilization energies per carbon. Moreover, in the cluster 10, adsorption energy is -0.54 eV for the first adsorbed Li, and it is less as much as -1.09 eV for the second. According to Table 1, if the adsorbed Li is closer to edge, the adsorption energy amount would increase, and therefore these adsorb states would be more useful for technological purposes. Consequently, edges of graphene islands are preferred for Li doped Nano-scale devices.

Table 1. A summary of calculation results for disordered graphene by Li atom

| Cluster number | Chemical Name | Energy (eV) | Adsorption Energy (eV) | Charge Of Li Atom(s) | Gap(eV) HOMO-LUMO | h (Å)[2] | Curvature (Å)[3] | Dipole Momentum (Debye) |
|---|---|---|---|---|---|---|---|---|
| 1 | Li $C_{24}H_{12}$ | -25290.00 | -0.00 | 0.40 | α:1.17 β:3.70 | 1.77 | -0.04 | 4.17 |
| 2 | Li $C_{24}H_{12}$ | -25290.28 | -0.27 | 0.33 | α:1.29 β:3.76 | 1.72 | -0.00 | 5.05 |
| 3 | Li $C_{54}H_{18}$ | -56503.38 | -0.54 | 0.44 | α:0.81 β:2.63 | 1.75 | -0.04 | 4.20 |
| 4 | Li $C_{54}H_{18}$ | -56503.38 | -0.54 | 0.43 | α:0.84 β:2.62 | 1.75 | 0.02 | 5.08 |
| 5 | Li $C_{92}H_{24}$ | -100162.99 | -0.54 | 0.46 | α:0.61 β:1.91 | 1.78 | -0.03 | 4.20 |
| 6 | Li $C_{92}H_{24}$ | -100162.97 | -0.54 | 0.45 | α:0.64 β:1.93 | 1.77 | -0.03 | 4.72 |
| 7 | Li $C_{92}H_{24}$ | -100162.99 | -0.82 | 0.44 | α:0.69 β:1.98 | 1.76 | -0.02 | 6.22 |
| 8 | Li $C_{92}H_{24}$ | -100163.26 | -0.82 | 0.36 | α:0.88 β:2.02 | 1.74 | -0.00 | 8.38 |
| 9 | Li $C_{92}H_{24}$ | -100163.26 | -0.82 | 0.39 | α:0.81 β:2.01 | 1.75 | -0.00 | 7.42 |
| 10 | $Li_2C_{92}H_{24}$ | -100366.80 | -1.09 | 0.50 | 0.53 | 1.78 | 0.00 | 0.00 |

The distributed charge of Li atom according to Mulliken population analysis is 40% for the first cluster and 33% for the second. So, the Li atom is less polarized for the cluster with closer Li to its edge (Table 1). For the cluster number 4 through 10 distributed charge of Li atom on the graphene surface are: 43%, 46%, 45%, 44%, 36%, 39% and 50%, respectively. The higher number of adsorbed Li on the graphene surface results in higher charge distributed on its surface; on the other hand, the minimum charge distribution belongs to the cluster number 2 and 8. So, it can be concluded that the neighborhood of Li atoms by hydrogen boundary edge is an important factor affecting charge distribution. As a result, the cluster number 2 and 8 with two covalence bonds and two hydrogen atoms have the lowest charge, compared with similar clusters. Cluster number 5 is not at the edge, but in the center, so it has the highest charge.

---

[2] Distance between Li(s) and the plane passes nearest benzene ring

[3] Distance between two surfaces, one passes central benzene ring and the other, passes hydrogen atoms, calculated by MERCURY 1.4.2 software.

As lithium has an unpaired electron, leading to a difference in spin-up and spin-down, when two lithium atoms are adsorbed simultaneously electrons get paired and magnet moment disappears. Consequently, all doped clusters, except cluster number 10, have spin-polarized energy levels (FIG.2). As a result, spin polarized cluster has a gap which size depends on adsorbed electron spin polarization, and the spin-down gap (alpha gap) is, 1eV, smaller than spin-up electron gap (beta gap) and there are dissimilar Fermi level energies for both up and down electron spins. Increasing size leads to decreased gap in graphene. However, both alpha and beta gaps decrease, when the size scale increases. Furthermore, gap of the cluster doped by two Li atoms is 0.5eV that decreases the gap of the graphene by 1.59eV [41] (Table 1).

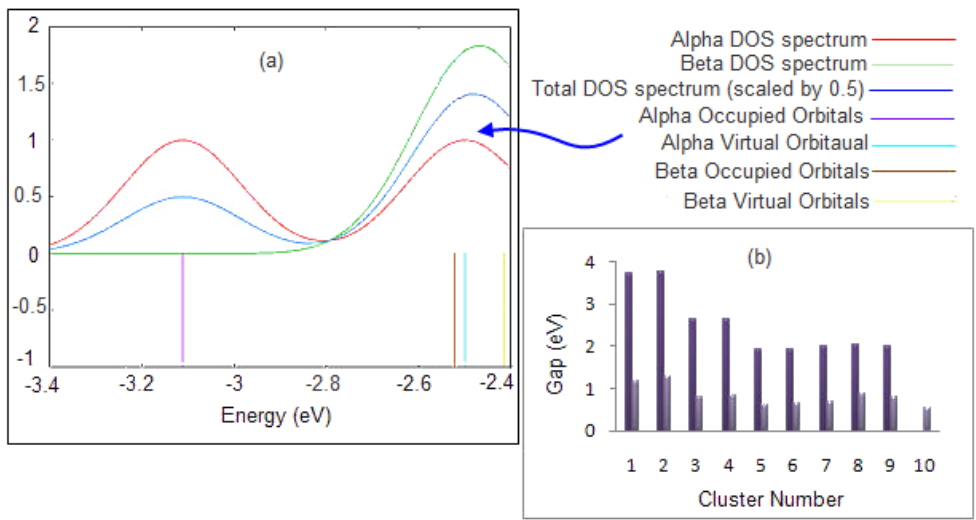

FIG.2. (a) DOS diagram for cluster number five. Electron states are spin polarized and the Fermi level for up and down spin are distinctive.(b) Alpha (lighter color) and beta (darker color) gap for graphene doped Li. This type of the material's electronic gap strongly depends on spin polarization.

It was observed that during the absorption process, two Li atoms are adsorbed in the center of their closest benzene ring, at the same distance from the cluster surface. The distance of both lithium atoms from the graphene surface are equal with an accuracy of $10^{-2}$ Å. Also, the results showed that higher charged lithium is closer to the cluster surface. If there are two lithium atoms adsorbed on graphene, they would be farther from the graphene in comparison to the case where one Li is adsorbed on the same cluster, while all these variations are in a range of 1.7 to 1.8 Å.

The Dipole momentum depends on the distance of Li from each cluster center. Therefore, when adsorbed Li distance from the central benzene ring increases, the graphene sheet will be more polarized. For the last cluster with two Li atoms on its surface, the dipole momentum is zero (by order of $10^{-2}$ Debye, Table 1).

Moreover, curvature is an important factor for Li absorption. As it increases, the number of Li adsorbed on graphene would also increase. The clusters have negative and positive curvatures, but higher curvatures belong to clusters with one Li atom at central benzene ring (FIG.3). When the size increases, the effect of Li adsorbed on graphene will decrease. So, the structure will tend to flat with a good accuracy. According to FIG.3, captivated Li affects geometry structure, as gravity may affect a jumping ball on a trampoline. Hydrogen atoms have the same role as edges of the trampoline. Hence, for a large-size cluster, the influence of one or more Li atoms on cluster curvature will decrease. As a result, it is better to use smaller clusters because of their capability of storing more numbers of Li atoms per surface area.

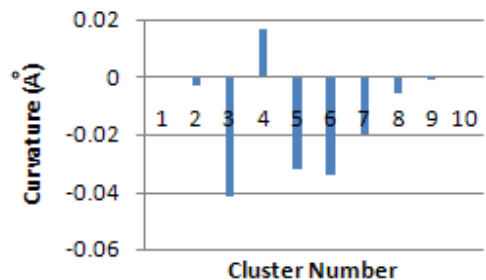

FIG.3: Curvature of graphene doped lithium. Cluster size, the number of adsorbed Li and the position of each from the edge are three determining factors for curvature.

**4. Improved Li ion Battery Suggestion**

The effect of temperature on adsorbed lithium, at room temperature (270<T<400K), is that the Li atoms will fall in boundary condition [25]. Although it is concluded from Table 1 that, because of the hydrogen boundary conditions, the distributed charge of Li cluster surface will decrease nearly by 10% on the edges, and this is an adverse factor for Li battery function. In addition, absorption of Li atoms on the edge decreases cluster curvature in compare with the adsorbing of one Li on its middle rings. According to more calculations, the sandwich structures' Li charge is logically relative to the single cluster's Li charge. Consequently, if a cluster has a larger Li charged, its sandwich structure will also have a larger Li charged. So, we can consider single graphene layers to judge about its sandwich structure charge storage capacity relative to another one. This suggestion could be used to avoid time consuming calculations which conclude the optimization for larger sandwich shape cases. Consequently, and based on structure's geometric symmetry, a Li battery structure as in FIG.4 is recommended.

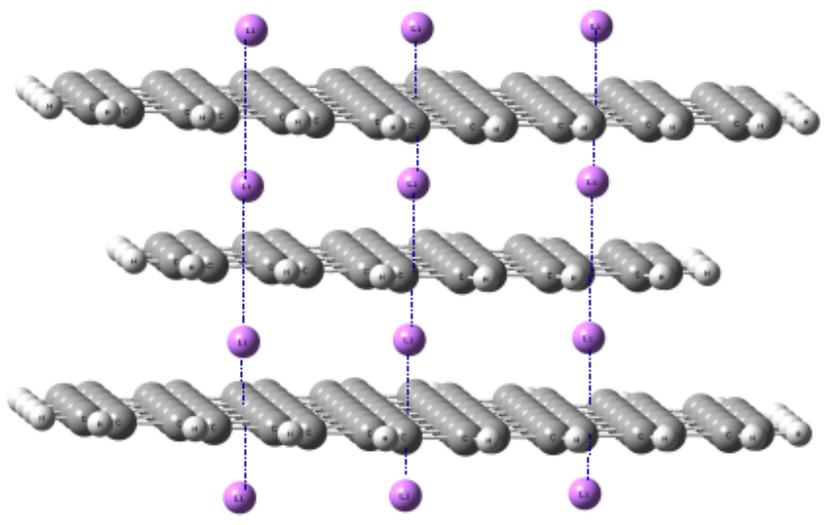

FIG.4. Suggested Li battery structure. According to this figure, two larger polycyclic clusters have a smaller cluster, and each has Li ions on both sides.

This structure has a smaller cluster between two larger clusters, and Li ions are adsorbed on both sides. We have checked this suggestion by use of a sandwich structure of $C_{24}H_{12}$ and $C_{54}H_{18}$ which has one Li atom between its two graphene surfaces. In comparison to a sandwich structure of only $C_{54}H_{18}$ this suggested structure store Li charge 20% more. Our calculation shows that the ideal distance between two adjacent clusters is 3.6 Å. According Table 1, higher number of Li atoms is captivated on a cluster more distant from the surface, and this is considered to be 3.6 Å. Moreover, research works on Nano graphene sheets, with a large enough interlayer spacing induced by using a carbon nanotube or $C_{60}$ spacer approach showed that high Li storage capacities make this 3D architecture possible [43]. This specific structure provides more wrapped surface in comparison to previously suggested structures [44], because the lower cluster elevates Li storage capability by increasing curvature factor. As we have discussed, all of Li atoms should be adsorbed in the center of a benzene ring. So, there is a preferred direction, and imaginary lines pass through three concentric benzene rings cutting surfaces vertically, as shown in FIG.4. Moreover, gravity effect is negligible for these cluster rings. Hence, it is predicted that movement for both graphene cluster reach their stable state i.e. when graphene surfaces are parallel. This result could also be useful for molecular machine designers. Finally, adsorbed Li on the edge, discussed in the previous section, increases adsorption energy amount, so it is preferred for lithium batteries, and molecular machine design.

## 5. Conclusion

Adsorption energy of graphene depends on cluster size, and the distance between Li atom and the center of structure. Also, there is not a meaningful relationship between adsorption energy and Li charge; however further distance from the cluster surface led to higher Li charge. For the cluster with one adsorbed Li, the minimum electrostatic potential occurs on the side opposed to the Li-adsorbed surface. Therefore, to improve Li function, it is suggested that the same number of Li ions should be exposed to both sides of graphene sheets. An important result for electronic industry is that cluster gap depends on both Li distance from the cluster center and cluster size. All of clusters doped by one Li are spin polarized and alpha gap decreases with cluster size, so does beta gap; in addition, we predict similar results for graphene adsorbed by an even number of Li atoms. Graphene cluster doped by two Li atoms is not spin polarized as it was predicted for an odd number of adsorbed Li on graphene clusters. Based on aforementioned results, an improved Li-ion battery with a smaller cluster between each two larger clusters is suggested. In addition, a movement is predicted for each graphene cluster to reach the stationary state, and make parallel graphene surfaces. This result could also be useful for molecular machine designers.

**Acknowledge**

We acknowledge useful discussions with Prof. Paul Schleyer, university of Georgia (USA), and Prof. Chandre Dharma-Wardana, University de Montréal (Canada). This work was made possible by the facilities of Computational Nanotechnology Supercomputing Center, Institute for Research in Fundamental Science (IPM), and National High Performance Computing Center, Isfahan University of Technology.

**References:**

[1] S.L.Cheekatia, Y.Xing, Y.Zhuang, H.Huang, Lithium storage characteristics for nano-graphene plates. In: Material challenges in alternative and renewable energy: Ceramic transaction, George Wicks G, Simon J, Zidan R, Lara-Curzio E, Adams T, Zayas J, et al: editor , Wiley, 224, 2011, pp. 117-27.

[2] K.Sato, M.Noguchi, A.Demachi, N.Oki, M.Endo, A mechanism of lithium storage in disordered carbons, Science. 264 (1994) 556-8.

[3] J.M.Tarascon, M.Armand, Issues and challenges facing rechargeable lithium batteries, Nature. 414 (2001) 359-67.

[4] P.Lian, X.Zhu, S.Liang, Z.Li, W.Yang, H.Wang, Large reversible capacity of high quality graphene sheets as an anode material for lithium-ion batteries, ElectrochimActa. 55 (2010) 3909-14.

[5] Y.Idota, T.Kubota, A.Matsufuji, Y.Maekawa, T.Miyasaka, Glasses for lithium batteries, Science. 276 (1997) 1309.

[6] P.Poizot, S.Laruelle, S.Grugeon, L.Dupont, J.M.Tarascon, Nano-sized transition-metal oxides as negative-electrode materials for lithium-ion batteries, Nature. 407 (2000) 496-9.

[7] H.Zhou, S.Zhu, M.Hibino, I.Honma, M.Ichihara, Lithium Storage in Ordered Mesoporous Carbon (CMK-3) with High Reversible Specific Energy Capacity and Good Cycling Performance, Adv. Mater. 15 (2003) 2107–11.

[8] P.L.Taberna, S.Mitra, P.Poizot, P.Simon, J.M.Tarascon, High rate capabilities $Fe_3O_4$-based Cu nano-architectured electrodes for lithium-ion battery applications, Nat. Mater. 5 (2006) 567-73.

[9] C.K.Chan, H.Peng, G.Liu, K.M.Wrath, X.F.Zhang, R.A.Huggins, et al, High-performance lithium battery anodes using silicon nanowires, Nat. Nanotechnol. 3 (2008) 31-5.

[10] K.S.Novoselov, A.K.Geim, S.V.Morozov, Y.Zhang, S.V.Dubonos, et al, Electric field effect in atomically thin carbon films, Science. 306 (2004) 666–9.

[11] J.C.Meyer, A.K.Geim, M.I.Katsnelson, K.S.Novoselov, T.J.Booth, S.Roth, et al, The structure of suspended graphene sheets, Nature. 446 (2007) 60–63.

[12] J.Hou, Y.Shao, M.Ellis, B.R.Moore, B.Yi, Graphene-based electrochemical energy conversion and storage: fuel cells, supercapacitors and lithium ion batteries, PCCP. 13 (2011) 15384–402.

[13] X.Du, I.Skachko, A.Barker, E.Y.Andrei, Approaching ballistic transport in suspended graphene, Nat.Nanotechnol. 3 (2008) 491–5.

[14] C.N.R.Rao, AK.Sood, R.Voggu, SubrahmanyamKS.Some novel attributes of graphene, J. Phys. Chem. Lett. 1 (2010) 572-80.

[15] G.Wang, X.Shen, J.Yao, J.Park, Graphene nanosheets for enhanced lithium storage in lithium ion batteries, Carbon. 47 (2009) 2049-53.

[16] G.Forte, F.Grossi, G.M.Lombardo, A.La Magna, G.G.N.Angilella, P.Pucci, et al, Modeling vacancies and hydrogen impurities in graphene: A molecular point of view, Phys. Lett. A. 372 (2008) 6168-74.

[17] C.M.Dharma-Wandra, M.Z.Zgierski, Magnetism and structure at vacant lattice sites in graphene, Physica E. 41 (2008) 80-3.

[18] D.Moran, F.Sthal, H.F.Bettinger, P.Schleyer, Towards Graphite: Magnetic Properties of Large Polybenzenoid Hydrocarbons, J. A. Chem. Soc. 125 (2003) 6746-52.

[19] J.Cho, S.Lim, J.Cha, N.Park, Analysis of the strong propensity for the delocalized diamagnetic π electronic structure of hydrogenated graphenes, Carbon. 49 (2011) 2665-70.

[20] E.J.Yoo, J.Kim, E.Hosono, H.S.Zhou, T.Kudo, I.Honma, Large Reversible Li Storage of Graphene Nanosheet Families for Use in Rechargeable Lithium Ion Batteries, Nano Lett. 8 (2008) 2277–82.

[21] G.Wang, B.Wang, X.Wang, J.Park, S.Dou, H.Ahn, K.Kim, Sn/graphene nanocomposite with 3D architecture for enhanced reversible lithium storage in lithium ion batteries, J. Mater. Chem. 19 (2009) 8378-84.

[22] N.Kurita, M.Endo, Molecular orbital calculations on electronic and Li-adsorption properties of sulfur-, phosphorus- and silicon-substituted disordered carbons, Carbon. 40 (2002) 253-60.

[23] J.R.Dahn, T.Zheng, Y.H.Liu, J.S.Xue, Mechanism for lithium insertion in carbonaceous materials, Science. 270 (1995) 590–3.

[24] Y.H.Liu, J.S.Xue, T.Zheng, J.R.Dahn, Mechanism of lithium insertion in hard carbons prepared by pyrolysis of epoxy resins, Carbon. 34 (1996) 193–200.

[25] H.Tachikawa, A.Shimizu, Diffusion dynamics of the li atom on amorphous carbon: A direct molecular orbital-molecular dynamics study, J. Phys. Chem. B. 110 (2005) 20445-50.

[26] C.A.Lin, T.Liedl, R.A.Sperling, M.T.Fernández-Argüelles, J.M.Costa-Fernández, R.Pereiro, Bioanalytics and biolabeling with semiconductor nanoparticles (quantum dots), J. Mater. Chem. 17 (2007) 1343-46.

[27] D.Wang, R.Kou, D.Choi, Z.Yang, Z.Nie, J.Li, et al, Ternary self-assembly of ordered metal oxide-graphene nanocomposites for electrochemical energy storage, ACS NANO. 4 (2010) 1587-95.

[28] M.Winter, R.J.Broad, What Are Batteries, Fuel Cells, and Supercapacitors?, Chem. Rev. 105 (2005) 1021.

[29] P.G.Brouce, B.Scrosati, J.M.Tarascon, Nanomaterials for rechargeable lithium batteries, Angew chem. IntEd. 47(2008) 2930-46.

[30] M.Walkihara, Recent developments in lithium ion batteries, Mater. Sci. Eng. R. 33 (2001) 109-34.

[31] A.Marquez, A.Vargas, P.B.Balbuena, Computational studies of lithium intercalation in model graphite in the presence of tetrahydrofuran, J. Electrochem. Soc. 145 (1998) 3328-34.

[32] M.Nakadaira, R.Saito, T.Kimura, G.Dresselhaus, M.S.Dresselhaus, Excess Li ions in a small graphite cluster, J. Mater. Res. 12 (1997) 1367-75.

[33] M.Fujita, K.Wakabayashi, K.Nakada, K.Kusakabe, Peculiar localized state at zigzag graphite edge, J. Phys.Soc. Jpn. 65 (1996) 1920-3.

[34] R.C.Boehm, A.Banerjee, Theoretical study of lithium intercalated graphite, J. Chem. Phys. 96 (1992) 1150-8.

[35] D.J.Hankinson, J.Almlöf, Cluster models for lithium intercalated graphite: electronic structures and energetic, J. Mol. Struct, THEOCHEM. 388 (1996) 245-56.

[36] S.Ishikawa, G.Madjarova, T.Yamaba, First-Principles Study of the Lithium Interaction with Polycyclic Aromatic Hydrocarbons, J. Phys. Chem. B. 105 (2001) 11986-93.

[37] H.Tachikawa, Y.Nagoya, T.Fukuzumi, Density functional theory (DFT) study on the effects of Li$^+$doping on electronic states of graphene, J. Power Sources. 195 (2010) 6148-52.

[38] M.J.Frisch, G.W.Trucks, H.B.Schlegel, G.E.Scuseria, M.A.RobbA, J.R.Cheeseman, et al, Gaussian 03, Revision B.02, Gaussian, Inc. Pittsburgh PA, 2003.

[39] A.D.Becke, Density-functional thermochemistry. III. The role of exact exchange, J. Chem. Phys. 98 (1993) 5648-52.

[40] C.Lee, W.Yang, R.G.Parr, Development of the Colle-Salvetti correlation-energy formula into a functional of the electron density, Phys. Rev. B. 37 (1998) 785-9.

[41] GaussSum 2.1 (c) 2007, Noel O'Boyl.

[42] N.Kheirabadi, A.Shafiekhani, M.A.Vesaghi, The ground state of quasi-circular graphene and graphene with vacancies, arXiv: 1004.2963v1 [cond-mat.mes-hall], 2010.

[43] E.Yoo, J.Kim, E.Hosono, H.S.Zhou, T.Kudo, I.Honma, Large reversible Li storage of graphene nanosheet families for use in rechargeable lithium ion batteries, Nano Lett. 8 (2008) 2277-82.

[44] T.Suzuki, T.Hasegawa, S.Mukai, H.A.Tamon, A theoretical study on storage states of Li ions in carbon anodes of Li ion batteries using molecular orbital calculations, Carbon. 41 (2003) 1933–39.